\documentstyle[12pt,axodraw]{article}
\newcommand{\be}{\begin{equation}}
\newcommand{\ee}{\end{equation}}
\newcommand{\bea}{\begin{eqnarray}}
\newcommand{\eea}{\end{eqnarray}}
\newcommand{\nn}{\nonumber}

\begin{document}
\title{A Dynamical Approach to Gauge Fixing}
\author{F. Loran\thanks{e-mail:
loran@cc.iut.ac.ir}\\ \\
  {\it Department of  Physics, Isfahan University of Technology (IUT)}\\
{\it Isfahan,  Iran,} \\
  {\it Institute for Studies in Theoretical Physics and Mathematics (IPM)}\\
{\it P. O. Box: 19395-5531, Tehran, Iran.}}
\date{}
\maketitle

\begin{abstract}We study gauge fixing in
the generalized Gupta-Bleuler quantization. In this method physical states are
defined to be simultaneous null eigenstates of a set of quantum invariants. We apply
the method to a solvable model proposed by Friedberg, Lee, Pang and Ren and show that
no Gribov-type copies appears by construction.
\end{abstract}
Gauge fixing is significant in quantization of gauge theories. Gribov ambiguities and
specially Gribov-type copies \cite{Gribov} have been a challenge for standard methods
of quantization. There are some methods to avoid Gribov ambiguities. For example
Gribov proposed to restrict field configurations only to those with positive
Faddeev-Popov determinant. Recently Klauder \cite{Kla} has formulated a new method of
quantization based on projection operators into gauge invariant states that avoids
any gauge fixing procedure and consequently Gribov ambiguities. In this article we
propose an alternative approach to this problem. Using the concept of quantum
invariants \cite{Ries}, we show that one can avoid Gribov ambiguities by quantizing
the whole phase space and then determining the physical (reduced) Hilbert space
instead of constructing the Hilbert space based on the reduced phase space
\cite{Self}.
\par Considering a gauge model given by a Hamiltonian $H(q,p)$ and a set of first
class constraints, there exist a generator of gauge transformation that satisfies the
following conditions \cite{Pons1,Pons2}, \bea
\{G,\phi^0_\mu\}|_{M_0}&=&0,\nn\\
\left(\{G,H\}+\frac{\partial}{\partial t}G\right)_{M_0}&=&0,\nn \eea where $\{\ ,\
\}$ stands for the Poisson brackets and $M_0$ is the submanifold locally defined by
primary (first class) constraints $\phi^0_\mu$. Gauge transformations $\delta
q=\{q,G\}$ and $\delta p=\{p,G\}$, determine gauge orbits i.e. gauge orbits are the
integral curves of the generator of gauge transformation. All points on a gauge orbit
correspond to physically equivalent configurations of the system. To eliminate gauge
freedom (gauge fixing) one may introduce gauge fixing conditions $\chi_i=0$ that
intersect the gauge orbits. A necessary condition for each $\chi_i$ to intersect the
gauge orbits is that it should not belong to the set of integral curves of $G$ i.e.
$\{\chi,G\}\neq 0$ \cite{Henbook,Gov1,Shi}. As is shown formally in Fig.1, a gauge
fixing condition $\chi$ that satisfies the condition $\{\chi,G\}\neq 0$ may never
intersect some of the gauge orbits or there may exist gauge equivalent copies. In
these cases, $\chi$'s do not fix (completely) the gauge and Gribov ambiguities emerge
\cite{Gribov}. Obviously, Gribov ambiguities would not emerge if one was able to
select a certain point on the gauge orbit by determining its coordinates. This can be
achieved by using the concept of invariants. It is well known \cite{Dirac} that the
classical dynamics of a system possessing gauge degrees of freedom is given by a
total Hamiltonian \be H_T=H+v_\mu\phi^0_\mu,\label{a4}\ee where $v_\mu$'s are
Lagrange multipliers. Therefore gauge degrees of freedom are those coordinates $q$'s
which their acceleration $\ddot{q}$ depend on undetermined Lagrange multipliers
$v(t)$. It can be easily verified that corresponding to each gauge degree of freedom
$q$ there exist a dynamical invariant $I_q$ defined by the relation \be
\frac{\partial}{\partial t}I_q+\{I_q,H_T\}=0.\label{inv1} \ee Consequently the
conditions $I_q=0$ eliminate gauge freedom by determining a point on the gauge orbit.
To do quantization one introduces the commutator algebra e.g.
$\{q,p\}\to\frac{1}{i}[\hat{q},\hat{p}]$ and defines physical states (reduced Hilbert
space) as the simultaneous null eigenstates of quantum invariants
$\hat{I}_i$\footnote{It is important to note that the eigenvalues of a quantum
invariant are independent of time and the particular solutions of the Schr\"{o}dinger
equation are different from the eigenstates of $\hat{I}_i$ only by time dependent
phase factors \cite{Mos}}. This method is introduced in our previous work and is
called the generalized Gupta-Bleuler quantization \cite{Self}. In general the form of
the Hamiltonian defined in the generalized Gupta-Bleuler quantization is not the same
as the operator form of the total Hamiltonian $H_T$. Fortunately in the case of
important models like Yang-Mills theory and the solvable model proposed by Friedberg,
Lee, Pang and Ren \cite{Fri} these two forms are the same.
\par
In the following we study the solvable model proposed by Friedberg, Lee, Pang and Ren.
This model can exhibit Gribov copies and is given by a $U(1)$ gauge invariant
Lagrangian \be
L=\frac{1}{2}\left[\left(\dot{x}+gy\xi\right)^2+\left(\dot{y}-gx\xi\right)^2+
\left(\dot{z}-\xi\right)^2\right]-V\left(\sqrt{x^2+y^2}\right),\label{b1}\ee or
equivalently in terms of polar coordinates \cite{Gov2} \be
L=\frac{1}{2}\dot{r}^2+\frac{1}{2}r^2\left(\dot{\theta}-g\xi\right)^2+
\frac{1}{2}\left(\dot{z}-\xi\right)^2-V(r).\label{b2}\ee The corresponding total
Hamiltonian is \be
H_T=\frac{1}{2}\left(p_r^2+\frac{L_z^2}{r^2}+p_z^2\right)+\xi\left(p_z+gL_z\right)+V(r)
+\ddot{\epsilon}(t)p_\xi,\label{b3}\ee where $p_r$, $L_z$, $p_z$ and $p_\xi$ are
momenta respectively conjugate to coordinates $r$, $\theta$, $z$, $\xi$ and
$\ddot{\epsilon}(t)$ is the Lagrange multiplier. Following the generalized
Gupta-Bleuler quantization the expectation value of the coordinate operators
$q_c=\left<\hat{q}\right>_{{\rm phys}}$ should satisfy the Euler-Lagrange equations
of motion obtained from the Lagrangian given in Eq.(\ref{b2}). This can be achieved
by introducing the Hamiltonian $\hat{H}^{(0)}=H_T(\hat{q},\hat{p})$, assuming the
commutator algebra $\{\ ,\ \}\to\frac{1}{i}[\ ,\ ]$ and the Schr\"{o}dinger equation
\be i\frac{\partial}{\partial t}\left|{\rm phys}\right>=\hat{H}^{(0)}\left|{\rm
phys}\right>.\label{c1}\ee In addition one should impose the condition
\be\left(\hat{p}_z+g\hat{L}_z\right)\left|{\rm phys}\right>=0\label{c11}\ee to
guarantee the validity of the Lagrangian constraint \bea
L_\xi\to\left((\dot{z}_c-\xi_c)+gr_c^2(\dot{\theta}_c-g\xi_c)\right)=0.\nn\eea Now for
convenience we identify $\left|{\rm phys}\right>$ as \bea \left|{\rm
phys}\right>=\left|\psi_\xi\right>\left|\psi_z\right>\left|\psi_\theta\right>
\left|\psi_r\right>.\label{c2}\eea It should be noted that our final result does not
depend on this particular form of $\left|{\rm phys}\right>$. One can easily verify
that the operator \be \hat{I}_\xi=\hat{\xi}-\dot{\epsilon}(t),\ee is an invariant.
This means that $\left|\psi_\xi\right>$ is the null eigenstate of $\hat{I}_\xi$: \bea
\hat{I}_\xi\left|\psi_\xi\right>=0,\nn\eea or equivalently
\be\left<p_\xi|\psi_\xi\right>=\exp\left(-ip_\xi\dot{\epsilon}(t)\right),\label{d1}\ee
where $\hat{p}_\xi\left|p_\xi\right>=p_\xi\left|p_\xi\right>$. Consequently, the
Scr\"{o}dinger equation Eq.(\ref{c1}) can be written as \be i\frac{\partial}{\partial
t}\left|\psi_z\right>\left|\psi_\theta\right>
\left|\psi_r\right>=\hat{H}^{(1)}\left|\psi_z\right>\left|\psi_\theta\right>
\left|\psi_r\right>,\label{d2}\ee where $\hat{H}^{(1)}$ is defined as follows, \bea
\hat{H}^{(1)}=\frac{1}{2}\left(\hat{p}_r^2+\frac{\hat{L}_z^2}{\hat{r}^2}+\hat{p}_z^2\right)+
\dot{\epsilon}(t)\left(\hat{p}_z+g\hat{L}_z\right)+V(\hat{r}).\nn\eea This Hamiltonian
leads to another quantum invariant $\hat{I}_z=\hat{z}-\epsilon(t)-\hat{p}_zt$. This
determines $\left|\psi_z\right>$ to be
\be\left<p_z|\psi_z\right>=\exp\left[-i\left(p_z\epsilon(t)+\frac{1}{2}p_z^2t\right)\right],
\label{d3}\ee and the Schr\"{o}dinger equation takes a new form \be
i\frac{\partial}{\partial t}\left|\psi_\theta\right>
\left|\psi_r\right>=\hat{H}^{(2)}\left|\psi_\theta\right>
\left|\psi_r\right>,\label{d4}\ee where $\hat{H}^{(2)}$ is \bea
\hat{H}^{(2)}=\frac{1}{2}\left(\hat{p}_r^2+\frac{\hat{L}_z^2}{\hat{r}^2}\right)+
g\dot{\epsilon}(t)\hat{L}_z+V(\hat{r}).\nn\eea Since $\hat{L}_z$ commutes with
$\hat{H}^{(2)}$ one can identify $\left|\psi_\theta\right>$ as eigenstates of
$\hat{L}_z$; \be
\left<L_z|\psi_\theta\right>=\exp\left(-igL_z\epsilon(t)\right),\label{e1}\ee where
$L_z$ is an integer due to periodicity condition on $\left|\psi_\theta\right>$.
Inserting Eq.(\ref{e1}) in Eq.(\ref{d4}) one obtains the true Schr\"{o}dinger
equation for $\left|\psi_r\right>$:\be i\frac{\partial}{\partial t}
\left|\psi_r^n\right>=\left[\frac{1}{2}\left(\hat{p}_r^2+\frac{n^2}{\hat{r}^2}\right)+
V(\hat{r})\right]\left|\psi_r^n\right>,\label{e2}\ee where $n=L_z$ is an integer.
Therefore the physical states are \be \left<p_\xi,p_z,L_z,r|{\rm phys}\right>=
\exp\left[-i\left(p_\xi\dot{\epsilon}(t)+p_z\epsilon(t)+\frac{1}{2}p_z^2t+
gL_z\epsilon(t)\right)\right]\psi^n(r). \label{f1}\ee It should be noted that the
condition given in Eq.(\ref{c11}) simply relates the eigenvalues of $\hat{p}_z$ and
$\hat{L}_z$ to each other i.e.
$p_z+gL_z=0$. Using Eq.(\ref{f1}), one verifies that \bea \xi_c&=&\dot{\epsilon}(t),\nn\\
z_c&=&\epsilon(t)+p_zt\nn\eea and \bea \theta_c=g\epsilon(t).\nn\eea This means that
gauge freedom is properly fixed with no ambiguity. This result can be compared with
that of refs.\cite{Fri,Gov2}. In one sense our method is similar to the method
defined in ref.\cite{Kla}; in both methods one introduces projection operators into
(physical) gauge invariant states. In principle this method can be applied to
Yang-Mills theory but first, one should find the clear form of invariants which of
course is not a straightforward task.
\section*{Acknowledgement}The author would like to
thank M. Haghighat and K. Samani for useful discussions and comments.
\newpage\ \vspace{3cm}\bea
\begin{picture}(400,0)(0,0) \SetScale{1}
\LongArrow(-20,-50)(-20,120)\LongArrow(-30,-40)(320,-40)
\Curve{(0,-20)(40,20)(80,100)}\Curve{(50,-20)(90,20)(130,-20)}
\Curve{(160,-10)(200,50)(240,20)(280,100)}\DashLine(-10,40)(300,40){3}
\Text(7,53)[0]{$\chi=0$} \Text(165,-65)[0]{Fig.1}
\end{picture}
\nn\eea


\newpage
\begin{thebibliography}{99}
\bibitem{Gribov} V. N. Gribov, Nucl. Phys. {\bf B139}, (1978) 1.
\bibitem{Kla} J. R. Klauder, Ann. Phys. {\bf 254}, (1997) 419; Nucl. Phys. {\bf
B547}, (1999) 397.
\bibitem{Ries} H. R. Lewis and W. B. Riesenfeld, J. Math. Phys. {\bf 10}, (1969) 1458.
\bibitem{Self} F. Loran, hep-th/0203117.
\bibitem{Pons1} C. Batlle, J. Gomis, X. Gracia and J. M. Pons, J.
Math. Phys. {\bf 30} (6), (1986) 1345.
\bibitem{Pons2} J. M. Pons and J. A. Garcia, Int. J. Mod. Phys. A {\bf 15}
(2000) 4681.
\bibitem{Henbook} M. Henneaux and C. Teitelboim
{\it "Quantization of Gauge System"} Princeton University Press, Princeton, New
Jersey, 1992.
\bibitem{Gov1} J. Govaerts, {\it Hamiltonian Quantization and
Constraines Dynamics}, Leuven Notes in Mathematical and Theoretical Physics, Vol. 4,
1991.
\bibitem{Shi} A. Shirzad and F. Loran, hep-th/9912289.
\bibitem{Dirac} P. A. M. Dirac, Can. J. Math. {\bf 2}, (1950) 129 ;
Proc. R. Soc. London Ser. A {\bf 246}, (1958) 326; {\it "Lectures on Quantum
Mechanics"} New York: Yeshiva University Press, 1964,
\bibitem{Mos} A. Mostafazadeh, {\it Dynamical Invariants, Adiabatic Approximation and
the Geometric Phase}, Nova Science Publishers, New York, 2001.
\bibitem{Fri} R. Friedberg, T. D Lee, Y. Pang and H. C. Ren, Ann. Phys. {\bf 246}, (1996)
381.
\bibitem{Gov2} V. M. Villanueva,
J. Govaerts, J-L. Lucio-Martinez, J.Phys. A{\bf 33}, (2000) 4183.
\end{thebibliography}
\end{document}